\documentclass{desyproc}

\usepackage{lineno}

\newcommand{\antibar}[1]{%
  \mkern 3mu \overline{\mkern -3mu #1 \mkern -0.5mu} \mkern 0.5mu%
}
\newcommand{\PPh}{\ensuremath{\gamma}}
\newcommand{\PvPh}{\ensuremath{\gamma^*}}
\newcommand{\PPom}{\ensuremath{\mathbb{P}}}
\renewcommand{\Pr}{\ensuremath{\rho}}
\newcommand{\Prz}{\ensuremath{\rho^0}}
\newcommand{\Pf}{\ensuremath{f_0}}

\newcommand{\Ppi}{\ensuremath{\pi}}

\newcommand{\Ppip}{\ensuremath{\pi^+}}
\newcommand{\Ppim}{\ensuremath{\pi^-}}
\newcommand{\Ppipm}{\ensuremath{\pi^\pm}}

\newcommand{\Pqq}{\ensuremath{q}}
\newcommand{\Pqqbar}{\ensuremath{\antibar{q}}}
\newcommand{\Pp}{\ensuremath{p}}
\newcommand{\Pn}{\ensuremath{n}}

\newcommand{\PAu}{\ensuremath{\text{Au}}}

\newcommand{\twopion}{\ensuremath{\Ppip\Ppim}}
\newcommand{\fourpion}{\ensuremath{\Ppip\Ppim\Ppip\Ppim}}

\newcommand{\lrbrk}[1]{{\left({#1}\right)}}
\newcommand{\lrBrk}[1]{{\left[{#1}\right]}}
\newcommand{\lrabs}[1]{{\left|{#1}\right|}}
\newcommand{\abs}[1]{{|{#1}|}}
\newcommand{\mean}[1]{{\langle{#1}\rangle}}

\newcommand{\orderOf}[1]{\ensuremath{\mathcal{O}\lrbrk{#1}}}
\newcommand{\D}[2][]{\operatorname{d^{#1} \mathnormal{#2}}}

\newcommand{\pT}{\ensuremath{p_T}}
\newcommand{\vpT}{\ensuremath{\vec{p}_T}}
\newcommand{\sqrtsnn}[1]{%
  \ensuremath{\sqrt{s_{_{\!N\!N}}}%
    \ifthenelse{\equal{#1}{}}%
    {}%
    { = #1~\text{GeV}%
    }
  }
}
\newcommand{\measresult}[4]{%
  \ensuremath{#1%
    \ifthenelse{\equal{#2}{}}%
    {}%
    {\pm #2%
      \ifthenelse{\equal{#3}{}}%
      {}%
      {_\text{stat.}}%
    }%
    \ifthenelse{\equal{#3}{}}%
    {}%
    {\pm #3_\text{syst.}}\text{#4}%
  }%
}

\newcommand{\mevc}{~\ensuremath{\text{MeV}\! / \!c}}

\newcommand{\gevcsq}{~\ensuremath{(\text{GeV}\! / \!c)^2}}
\newcommand{\mevcc}{~\ensuremath{\text{MeV}\! / \!c^2}}

\newcommand{\tenpow}[2][]{%
  \ifthenelse{\equal{#1}{}}
  {\ensuremath{10^{#2}}}
  {\ensuremath{{#1} \cdot 10^{#2}}}
}

\newcommand{\figref}[1]{{Fig.~\ref{fig:#1}}}
\newcommand{\Figref}[1]{{Figure~\ref{fig:#1}}}
\newcommand{\equref}[1]{{Eq.~\eqref{eq:#1}}}

\newcommand{\others}{\textit{et al.}}

\begin{document}

\title{Photoproduction in Ultra-Peripheral Relativistic Heavy Ion Collisions at STAR}

\author{{\slshape Boris Grube$^1$} for the STAR Collaboration \\[1ex]
$^1$Excellence Cluster Universe, Technische Universit\"at M\"unchen, Garching, Germany}

\contribID{64}

\desyproc{DESY-PROC-2009-xx}
\acronym{EDS'09} 
\doi  

\maketitle


\begin{abstract}
  In ultra-peripheral relativistic heavy ion collisions the beam ions
  scatter at impact parameters larger than the sum of their radii, so
  that they interact via long range electromagnetic forces. Due to the
  Lorentz-boost of the beam particles, the exchanged virtual photons
  have high energies and can induce the photoproduction of
  vector-mesons. We present recent results of the STAR experiment at
  RHIC on $\Prz(770)$ production in \PAu-\PAu\ ultra-peripheral
  collisions at various energies. STAR has also observed the
  photoproduction of \fourpion, which is related to the still poorly
  known excited states of the \Prz.
\end{abstract}

\section{Introduction}

The electromagnetic field of a nucleus which is moving at relativistic
velocities can be approximated by a flux of quasi-real virtual photons
using the Weizs\"acker-Williams approach~\cite{weizsacker_williams}.
The number of photons scales with the atomic charge~$Z$ squared, so
that fast moving heavy nuclei create intense photon fluxes. Relativistic
heavy ions may thus be used as photon sources or targets.

In Ultra-Peripheral relativistic heavy ion Collisions (UPCs) the
long-range electromagnetic interactions are separated from the
otherwise indistinguishable hadronic interactions by requiring impact
parameters~$b$ larger than the sum of the nuclear radii~$R_A$ of the
beam ions. Due to the large Lorentz-boosts of the beam particles, it is possible
to study photonuclear reactions as well as photon-photon
interactions at high energies in UPCs~\cite{baur_krauss_bertulani_hencken}.

\begin{figure}[t]
  \vspace*{-1ex}
  \centering
  \includegraphics[height=0.2\textheight]{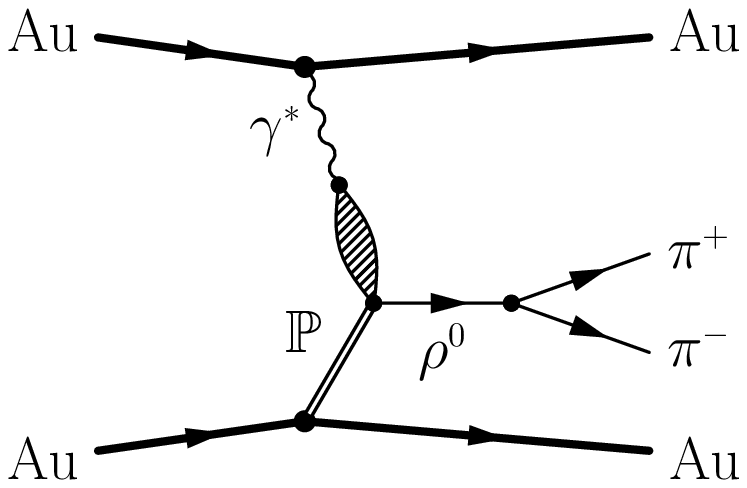} \qquad
  \includegraphics[height=0.2\textheight]{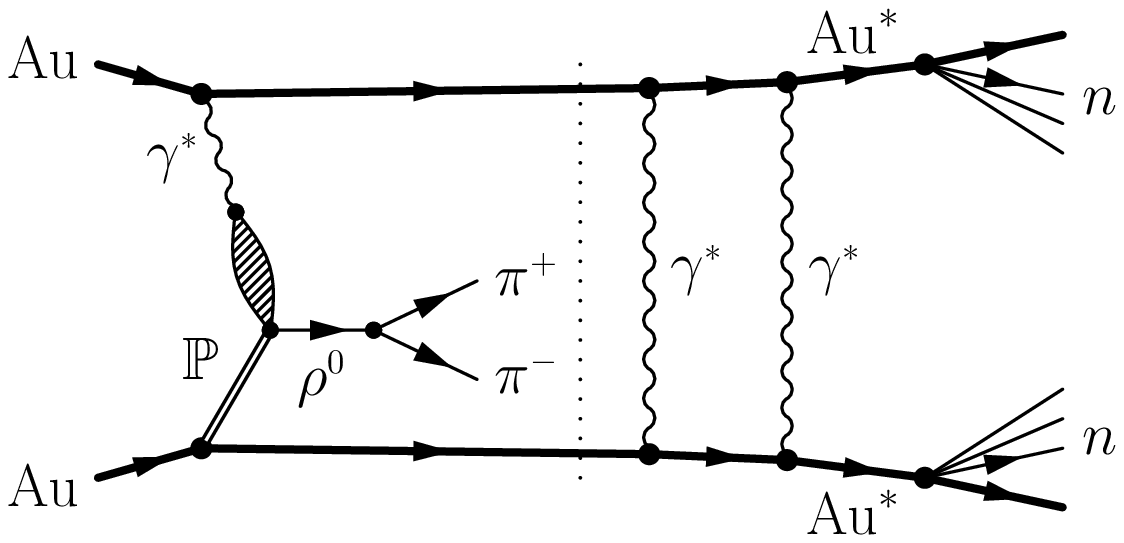} \\[-2ex]
  \raggedright
  \hspace{0.2\textwidth} \textbf{a)} \hspace{0.45\textwidth} \textbf{b)}
  \vspace*{-1ex}
  \caption{Schematic view of the photonuclear production of a
    $\Prz(770)$ meson in an ultra-peripheral \PAu-\PAu\ collision and
    its subsequent decay into two charged pions. The meson is produced
    in the fusion processes of a virtual photon \PvPh\ and a Pomeron
    \PPom. a)~shows the exclusive reaction, b)~the one with mutual
    Coulomb excitation of the beam ions and following neutron emission.}
  \label{fig:vectorMesonProd}
\end{figure}

The photoproduction of vector mesons is a typical process in UPCs. A
virtual photon, radiated by the ``emitter'' nucleus, fluctuates into a
\Pqq\Pqqbar~pair, which scatters elastically off the ``target''
nucleus and emerges as a real vector meson (cf.
\figref{vectorMesonProd}a). At high energies the scattering can be
described in terms of soft Pomeron exchange. The cross section is
strongly enhanced at low transverse momenta $\pT \lesssim 2 \hbar /
R_A$ of the produced meson, because the \Pqq\Pqqbar~pair couples
coherently to the entire nucleus. For these coherent processes the
cross section depends on the nuclear form factor $F(t)$, where $t$ is
the squared four-momentum transfer to the target nucleus. For larger
\pT\ the \Pqq\Pqqbar~pairs couple to the individual nucleons within
the target nucleus resulting in a smaller cross section which scales
approximately with the mass number~$A$ modulo corrections for the
nuclear absorption of the meson.

Because of the intense photon flux in the case of heavy beam ions, the
photoproduction of vector mesons may be accompanied by Coulomb
excitation of the beam particles. The excited ions decay mostly via
neutron emission~\cite{baltz_baur} which is a distinctive event
signature that can be utilized in the trigger decision. In lowest
order the vector meson photoproduction accompanied by mutual nuclear
dissociation of the beam ions is a three-photon process. One photon
produces the vector meson and two additional photons excite the nuclei
(see \figref{vectorMesonProd}b). In good approximation all three photon
exchanges are independent so that the cross section can be
factorized~\cite{baltz_baur}:

\begin{equation*}
  \sigma_{V,\, x\Pn\, x\Pn} =
    \int\!\D[2]{b} \lrBrk{1 - P_\text{had}(b)} \cdot P_V(b) \cdot P_{x\Pn, 1}(b) \cdot P_{x\Pn, 2}(b),
\end{equation*}

\noindent
where $P_\text{had}(b)$~is the probability for hadronic interaction,
$P_V(b)$~the probability to produce a vector meson~$V$\!\!, and
$P_{x\Pn, i}(b)$ the probability that nucleus~$i$ emits
$x$~neutrons. Compared to exclusive photonuclear vector meson
production, reactions with mutual Coulomb excitation have smaller
median impact parameters.

In this paper we present recent results from the STAR experiment at
the Relativistic Heavy Ion Collider (RHIC). The Solenoidal Tracker At
RHIC (STAR) uses a large cylindrical Time Projection
Chamber~(TPC)~\cite{star_tpc} with 2~m radius and 4.2~m length,
operated in a 0.5~T solenoidal magnetic field to reconstruct charged
tracks. For tracks with pseudorapidity $|\eta| < 1.2$ and transverse
momentum $\pT > 100\mevc$ the tracking efficiency is better than
85~\%. The UPC trigger is based on two detector systems: The two Zero
Degree Calorimeters (ZDCs)~\cite{star_zdc} which sit at $\pm 18$~m
from the interaction point and measure neutral particles emitted in
very forward direction. They have an acceptance close to unity for the
neutrons originating from nuclear dissociation of the beam ions. The
second trigger detector system used to select UPC events is the
Central Trigger Barrel (CTB)~\cite{star_ctb}. It is an array of 240
plastic scintillator slats that surrounds the TPC and provides
information about the charged-particle multiplicity.

Two basic types of trigger algorithms are used: The ``topology''
trigger requires a low overall charged-particle multiplicity and
subdivides the CTB into four azimuthal quadrants. Events with
coincident hits in the left and right quadrants are recorded thereby
selecting roughly back-to-back pion pairs. The top and bottom
quadrants are used to veto cosmic rays which otherwise could be
reconstructed as unlike-sign particle pairs with zero transverse
momentum and rapidity. Since there is no requirement on the energy
deposit in the ZDCs, the ``topology'' data mainly contain exclusively
produced vector mesons.  In contrast to this the ``minimum bias''
trigger selects UPC events, where both beam ions dissociated by
requiring coincident energy deposits in the ZDCs in addition to a low
total charged-particle multiplicity in the CTB.

In the offline analysis two- and four-prong events are selected by
requiring two and four charged tracks, respectively, in the TPC to
have zero net charge and to form a common (primary) vertex. All tracks
are assumed to be pions. In order to suppress backgrounds from
beam-gas interactions, peripheral hadronic interactions, and pile-up
events in addition a low overall charged-track multiplicity is
required. Backgrounds from pile-up events, beam-gas interactions, and
--- in the case of the two-prong sample --- cosmic rays are reduced by
selecting events with their primary vertex in a region close to the interaction
diamond. Cosmic ray backgrounds in the two-prong ``topology'' sample
are suppressed further by excluding events with rapidities $y_\Pr
\approx 0$. Due to the ZDC requirement in the ``minimum bias''
trigger, the cosmic ray background is nearly completely
removed. Finally, coherent events are selected by requiring small
transverse momenta of $\pT < 150$\mevc\ for the produced vector
mesons.

\mathversion{bold}
\section{Coherent Photoproduction of $\Prz(770)$}
\mathversion{normal}

There are at least three models that describe the production of
$\Prz(770)$ mesons in ultra-peripheral collisions: The model of Klein
and Nystrand~(KN)~\cite{KN} uses the Vector Dominance Model~(VDM) for
the virtual photon and a classical mechanical approach for the
scattering on the target nucleus, based on data from $\PPh\, \Pp \to
\Prz(770)\,\, \Pp$ experiments. The Frankfurt, Strikman, and
Zhalov~(FSZ) model~\cite{FSZ} employs a generalized VDM to describe
the virtual photon and a QCD Gribov-Glauber approach for the
scattering. The model of Gon\c{c}alves and Machado~(GM)~\cite{GM}
takes into account nuclear effects and parton saturation phenomena by
using a QCD color dipole approach.

The coherent cross section of $\Prz(770)$ production accompanied by
mutual nuclear dissociation of the beam ions $\sigma^\text{coh}_{\Pr,
  x\Pn x\Pn}$ is measured using ``minimum bias'' data. The \Prz~yield
is estimated by fitting the invariant mass peak of the acceptance
corrected $m_{\twopion}$ distribution and extrapolating the result
from the experimentally accessible rapidity range of $\abs{y_\Pr} < 1$
to the full solid angle using the KN~model~\cite{KN}. In \PAu-\PAu\
collisions at \sqrtsnn{200} the cross section was measured to be
$\sigma^\text{coh}_{\Pr, x\Pn x\Pn} =
\measresult{31.9}{1.5}{4.5}{mb}$; at \sqrtsnn{130} the value is
\measresult{28.3}{2.0}{6.3}{mb}~\cite{star_upc_rho}.

\begin{figure}[t]
  \vspace*{-1ex}
  \centering
  \includegraphics[height=0.26\textheight]{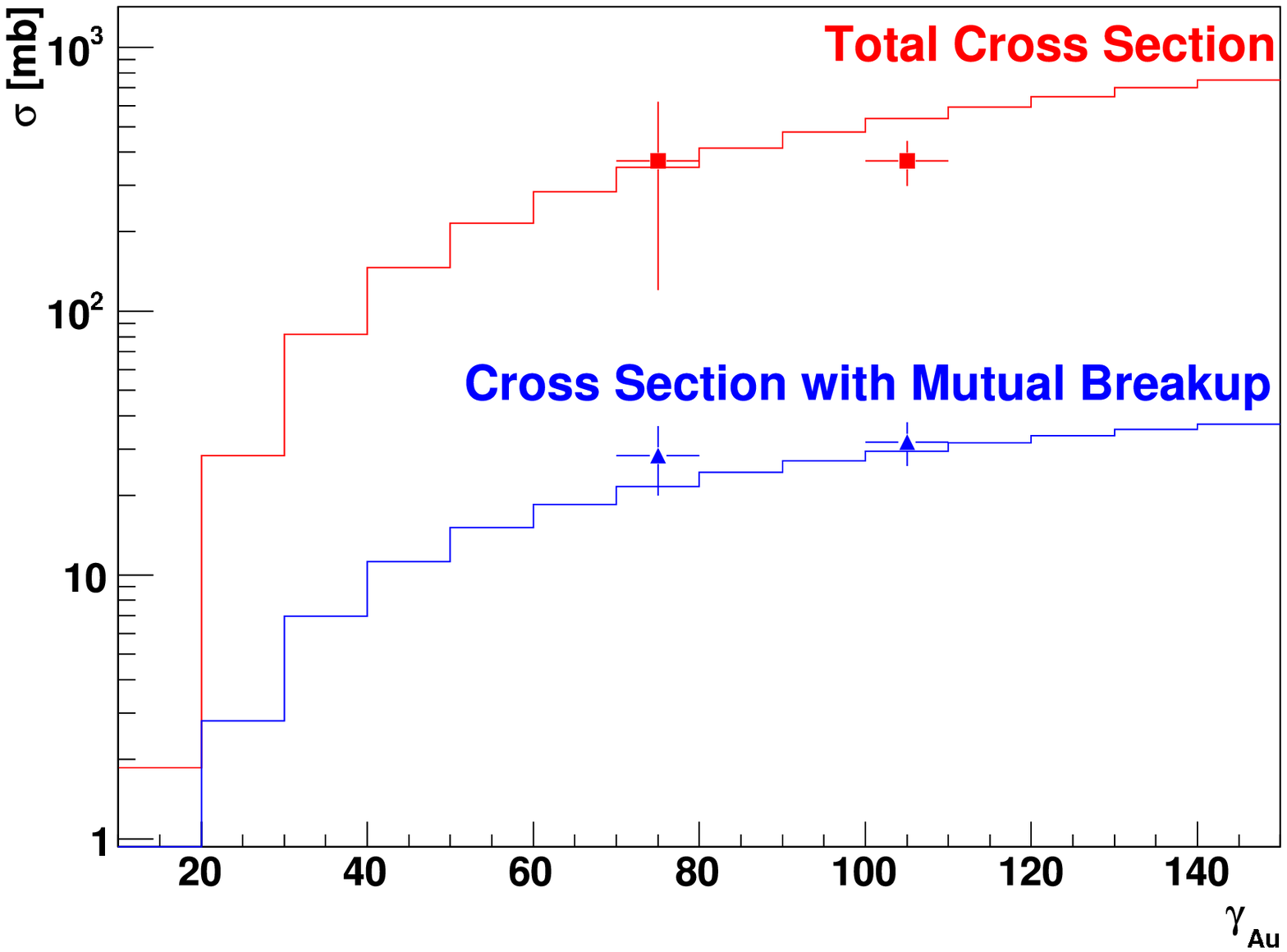} \hfill
  \raisebox{-1ex}{\includegraphics[height=0.29\textheight]{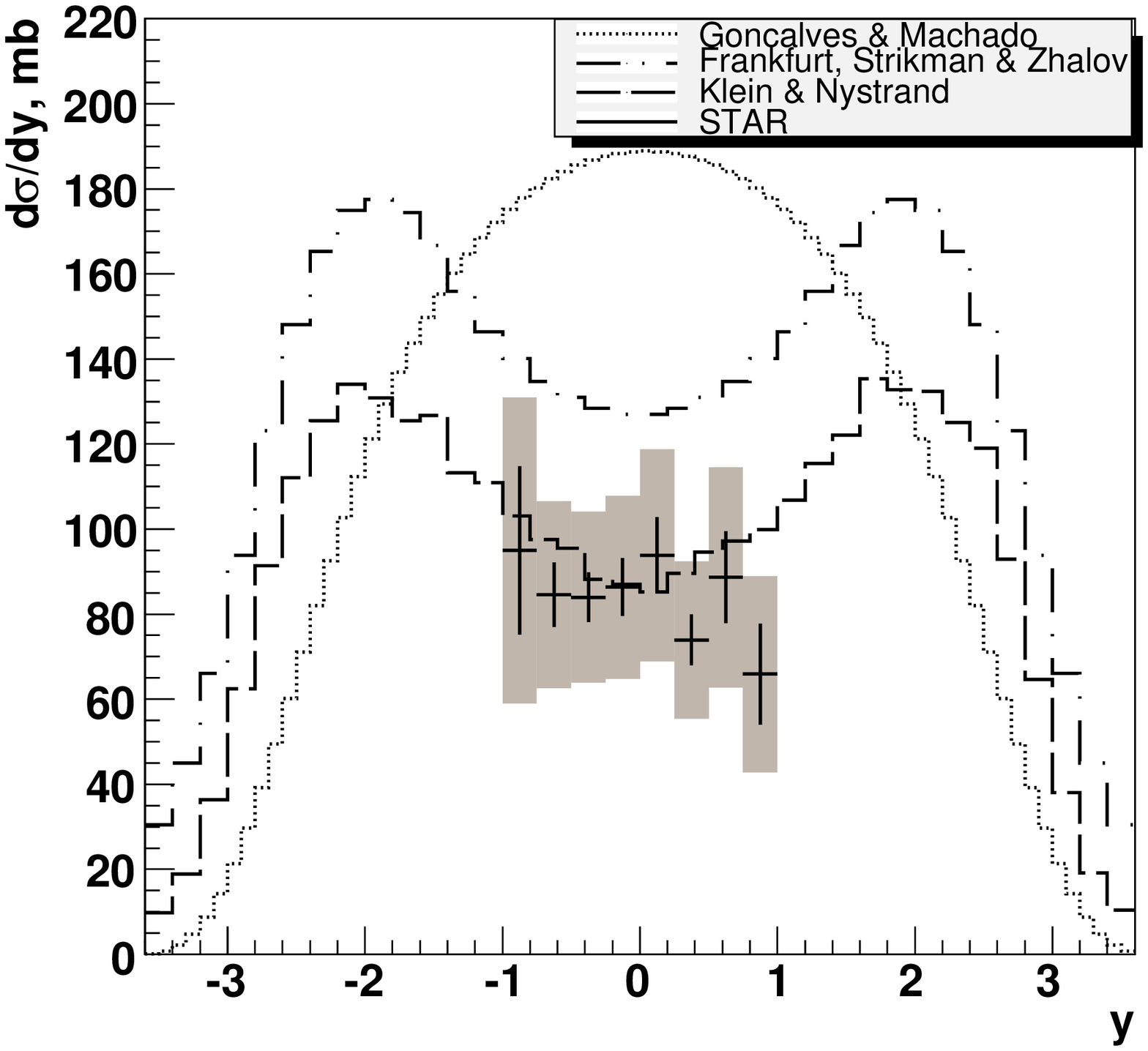}} \\
  \raggedright
  \hspace{0.24\textwidth} \textbf{a)} \hspace{0.50\textwidth} \textbf{b)}
  \vspace*{-1ex}
  \caption{a)~Energy dependence of the total coherent cross section
    (red) and the one with mutual nuclear dissociation (blue) in
    comparison to the KN~model predictions (continuous histogram).
    b)~Comparison of the measured total cross section for coherent
    \Prz~production with theoretical
    predictions~\cite{star_upc_rho}. The vertical line at each point
    shows the statistical error. The shaded area displays the sum of
    statistical and systematic errors. The dashed line represents the
    KN~\cite{KN}, the dash-dotted line the FSZ~\cite{FSZ}, and the
    dotted one the GM~model~\cite{GM}.}
  \label{fig:rhoCrossSect}
\end{figure}

Since the efficiency of the ``topology'' trigger is not well known,
the total cross section $\sigma^\text{coh}_{\Pr, \text{tot}}$ is
estimated by applying coherent cross section ratios for different
nuclear excitation states, which are extracted from the ``topology''
data, to the cross section values for mutual excitation. This way the
total coherent \Prz\ production cross section is measured to be
\measresult{530}{19}{57}{mb} at \sqrtsnn{200} and
\measresult{460}{220}{110}{mb} at
\sqrtsnn{130}~\cite{star_upc_rho}. In \figref{rhoCrossSect}a the
total cross sections and the cross sections with mutual nuclear
dissociation are compared to the KN~model predictions.

\Figref{rhoCrossSect}b shows the measured total coherent \Prz\
production cross section as a function of rapidity and compares to the
various model predictions. Due to the limited experimentally
accessible rapidity range of $\abs{y_\Pr} < 1$, it is not possible to
discriminate the models based on the shape of their rapidity
distribution. Considering the cross section values, the KN~model agrees
best with the data.

In \Prz~production an interesting interference phenomenon is caused by
the fact that the \Prz\ is produced close (\orderOf{1~\text{fm}}) to the
target nucleus and that the emitter and the target nucleus are
indistinguishable. Because the impact parameter is larger than the sum
of the nuclear radii of the projectiles, the system essentially acts
like a two-slit interferometer with slit separation
$\abs{\vec{b}}$. Either nucleus~A emits a virtual
photon which scatters off nucleus~B or vice versa. The two
indistinguishable processes are related by parity transformation and,
since the \Prz\ has negative intrinsic parity, the amplitudes have to
be subtracted~\cite{KN_int}:

\begin{equation*}
  \sigma(\vpT, \vec{b}, y_\Pr) = \lrabs{A(\pT, b, y_\Pr) - A(\pT, b, -y_\Pr)\, e^{i \vpT \cdot \vec{b}}}^2
\end{equation*}

\noindent
At mid-rapidity $A(\pT, b, y_\Pr) \approx A(\pT, b, -y_\Pr)$ so that the above equation simplifies to

\begin{equation*}
  \sigma(\vpT, \vec{b}, 0) = 2\, \lrabs{A(\pT, b, 0)}^2\, \lrBrk{1 - \cos(\vpT \cdot \vec{b})}
\end{equation*}

\begin{wrapfigure}[20]{r}{0.525\textwidth}
  \centering
  \includegraphics[width=0.525\textwidth]{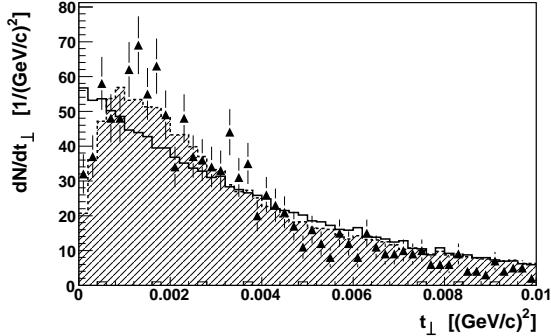} \\[-3ex]
  \caption{Uncorrected $t$ spectrum of \Prz\ in the rapidity range
    $\abs{y_\Pr} < 0.5$ for the ``minimum bias''
    data~\cite{star_upc_rho_int}. The points represent the data. The
    dashed (filled) histogram is a simulation that includes the
    interference effect, whereas the solid histogram is a simulation
    without interference.}
  \label{fig:rhoInt}
\end{wrapfigure}

\noindent
The interference is destructive for transverse momenta $\pT \lesssim
\hbar / \mean{b}$. \Figref{rhoInt} shows the $t (\approx \pT^2)$
distribution, which is roughly exponential at larger $t$, but has a
significant downturn for $t < 0.0015$\gevcsq, consistent with
the Monte-Carlo simulation that includes the interference effect.

The flight path $\beta \gamma c \tau$ of the produced \Prz\ is much
smaller than the impact parameter so that the \Prz\ decays at two
well-separated points in space-time. This means that the amplitudes
overlap and interfere only \emph{after} the decay and that the
interference must involve the \twopion\ final state. Interference is
only possible, if the final state wave function is entangled,
nonlocal, and not factorizable into individual \Ppipm\ wave functions.

The strength of the interference is extracted from the data by fitting
the $t$ distribution with the function

\begin{equation*}
  \frac{\D{N}}{\D{t}} = a\, e^{-kt}\, \lrBrk{1 + c\, (R(t) - 1)}
\end{equation*}

\noindent
where $k$ is the slope parameter and $c$ the spectral modification
parameter that measures the interference. A value of $c = 0$ would
correspond to no interference, a value of $c = 1$ to the interference
predicted by the KN~model~\cite{KN,KN_int}. The deviation of the $t$
distribution from the exponential shape due to the interference effect
is parameterized by the function $R(t)$ which is determined from the
ratio of the simulated $t$ spectrum with and without interference. The
measured spectral modification parameter of \measresult{87}{5}{8}{\%}
shows that the interference is significant~\cite{star_upc_rho_int},
which means that the \twopion\ final state wave function retains
amplitudes for all possible \Prz\ decays, long after the decay
occurred. The
system is thus an example of the Einstein-Podolsky-Rosen
paradox~\cite{EPR} with continuous variables momentum and position.

\mathversion{bold}
\section{Coherent Photoproduction of \fourpion\ Final States}
\mathversion{normal}

\begin{figure}[t]
  \vspace*{-1ex}
  \centering
  \includegraphics[width=0.49\textwidth]{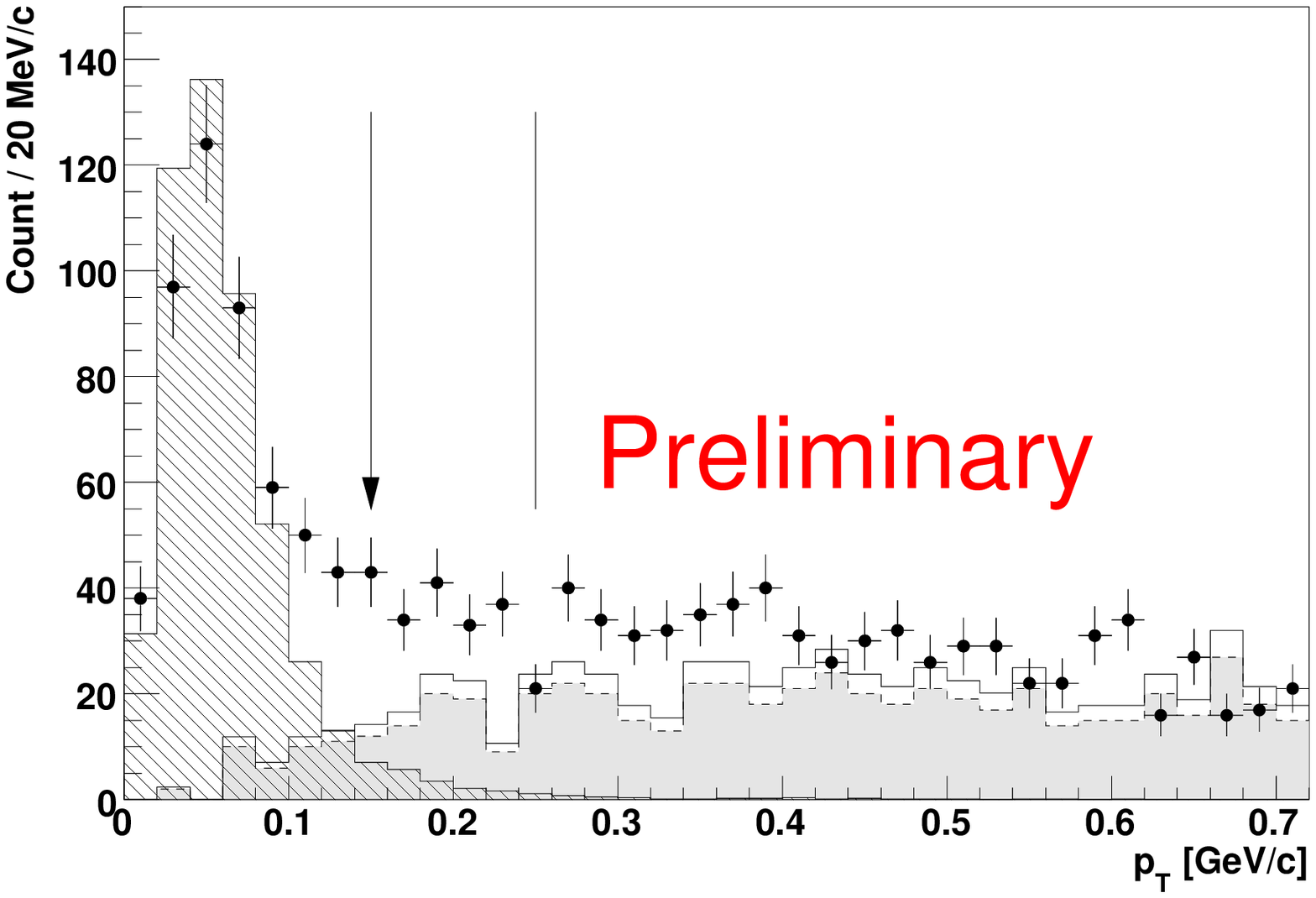} \hfill
  \includegraphics[width=0.49\textwidth]{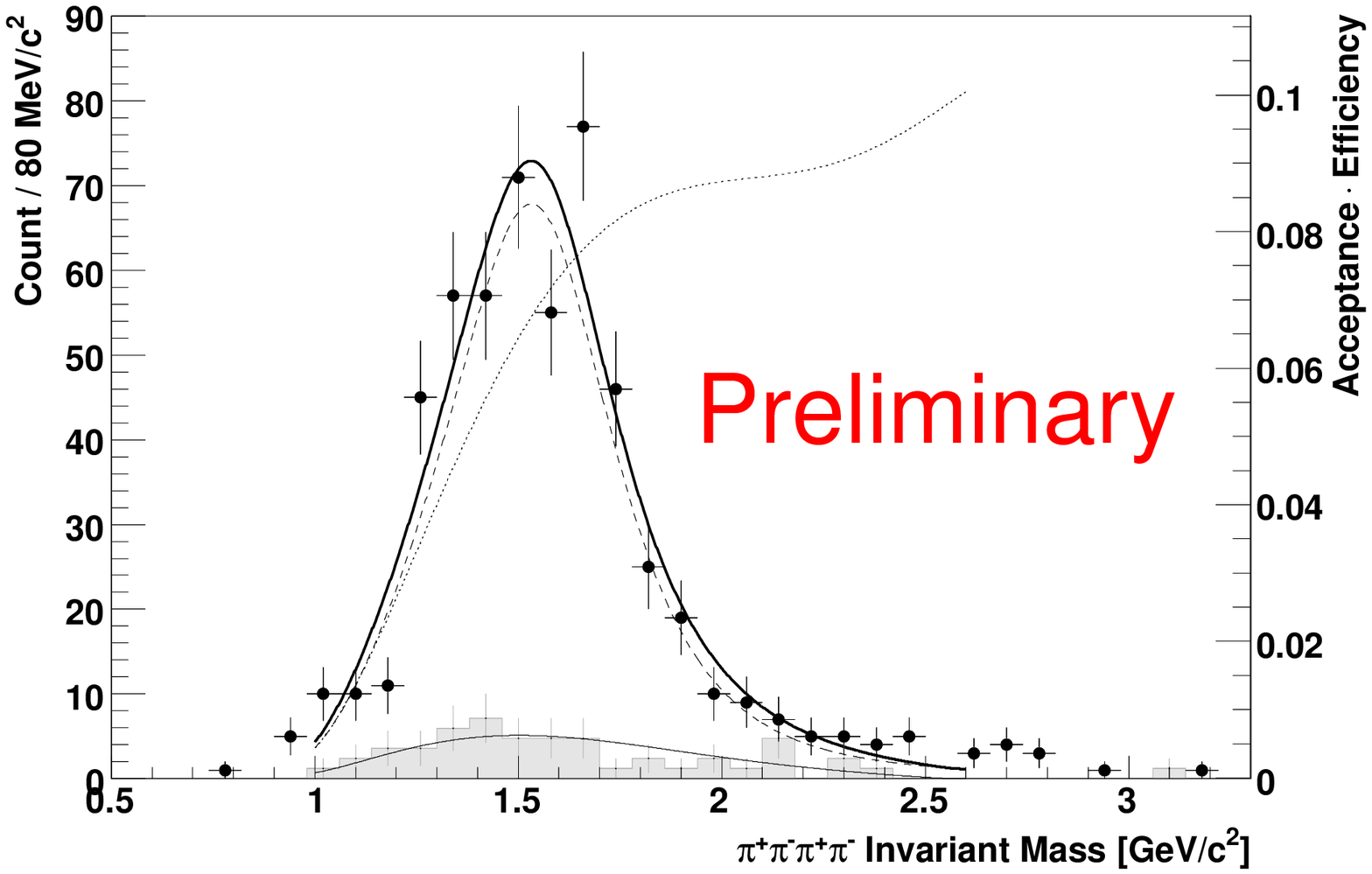} \\
  \raggedright
  \hspace{0.22\textwidth} \textbf{a)} \hspace{0.49\textwidth} \textbf{b)}
  \vspace*{-1ex}
  \caption{a)~\fourpion\ transverse momentum distribution: At low
    transverse momenta the four-prong couples coherently to the entire
    nucleus leading to a strong enhancement of the cross section. The
    hatched filled histogram shows the expected distribution from
    simulation. The background for the coherent part is estimated from
    $+2$ or $-2$ charged four-prong combinations by normalizing
    their \pT\ distribution (gray filled histogram) to that of the
    neutral four-prongs in the region of $\pT > 250\mevc$ (vertical
    line) yielding the unfilled histogram. b) Invariant mass
    distribution of coherently produced \fourpion: The points
    represent the data, the gray filled histogram is the background
    estimated from charged four-prongs. The thick black line shows the
    fit of the modified $S$-wave Breit-Wigner of \equref{rhoPrimeMass} on top of a second order
    polynomial background (thin black line) taking into account the
    detector acceptance in the region $\abs{y} < 1$ (rising
    dotted line). The dashed curve represents the signal
    curve without background.}
  \label{fig:rhoPrimePtMass}
\end{figure}

Coherent \fourpion\ production in ultra-peripheral collisions
accompanied by mutual nuclear dissociation of the beam ions was
measured in \PAu-\PAu\ collisions at \sqrtsnn{200} using the ``minimum
bias'' data. The transverse momentum spectrum of the neutral
four-prongs exhibits an enhancement at low \pT\, characteristic for
coherent production (cf. \figref{rhoPrimePtMass}a). The data show a
broad peak in the \fourpion\ invariant mass distribution (see
\figref{rhoPrimePtMass}b), similar to what was seen in earlier
fixed-target photoproduction
experiments~\cite{four_pion_photo_prod,four_pion_fnal}. This peak is
usually attributed to the excited \Prz~states $\Pr(1450)$ and
$\Pr(1700)$. However, the exact nature of these states is still
controversial.

The \fourpion\ invariant mass distribution was fitted with a $S$-wave
Breit-Wigner modified by a phenomenological Ross-Stodolsky factor~\cite{ross_stodolsky}:

\begin{equation}
  f(m) = A \cdot \lrbrk{\dfrac{m_0}{m}}^n\!\!\cdot \dfrac{m_0^2 \Gamma_0^2}{(m_0^2 - m^2)^2 + m_0^2 \Gamma_0^2} + f_\text{BG}(m)
  \label{eq:rhoPrimeMass}
\end{equation}

\noindent
The non-interfering background $f_\text{BG}$ was parameterized by a
second order polynomial which was extracted from the invariant mass
distribution of $+2$ or $-2$ charged four-prongs. Taking into account
the experimental acceptance, the fit yields a resonance mass of $1540
\pm 40$\mevcc\ and a width of $570 \pm 60$~MeV. The Ross-Stodolsky
exponent has a value of $n = 2.4 \pm 0.7$, however, mass and width
depend strongly on the value of $n$.

Using the acceptance-corrected \fourpion\ yield from the above fit and
the respective $\Prz(770)$ yield from the \twopion\ invariant mass
distribution the cross section ratio
$\sigma^\text{coh}_{4\Ppi, x\Pn x\Pn} / \sigma^\text{coh}_{\Pr, x\Pn
  x\Pn}$ is estimated to be $13.4 \pm 0.8$~\%, where again the
KN~model~\cite{KN} was used to extrapolate from the experimentally
accessible rapidity region $\abs{y} < 1$ to the full solid angle.
Using the measured coherent \Prz\ production cross section
$\sigma^\text{coh}_{\Pr, x\Pn, x\Pn}$ the \fourpion\ production cross
section is $4.3 \pm 0.3$~mb.

\begin{figure}[htb]
  \vspace*{-1ex}
  \centering
  \includegraphics[width=0.49\textwidth]{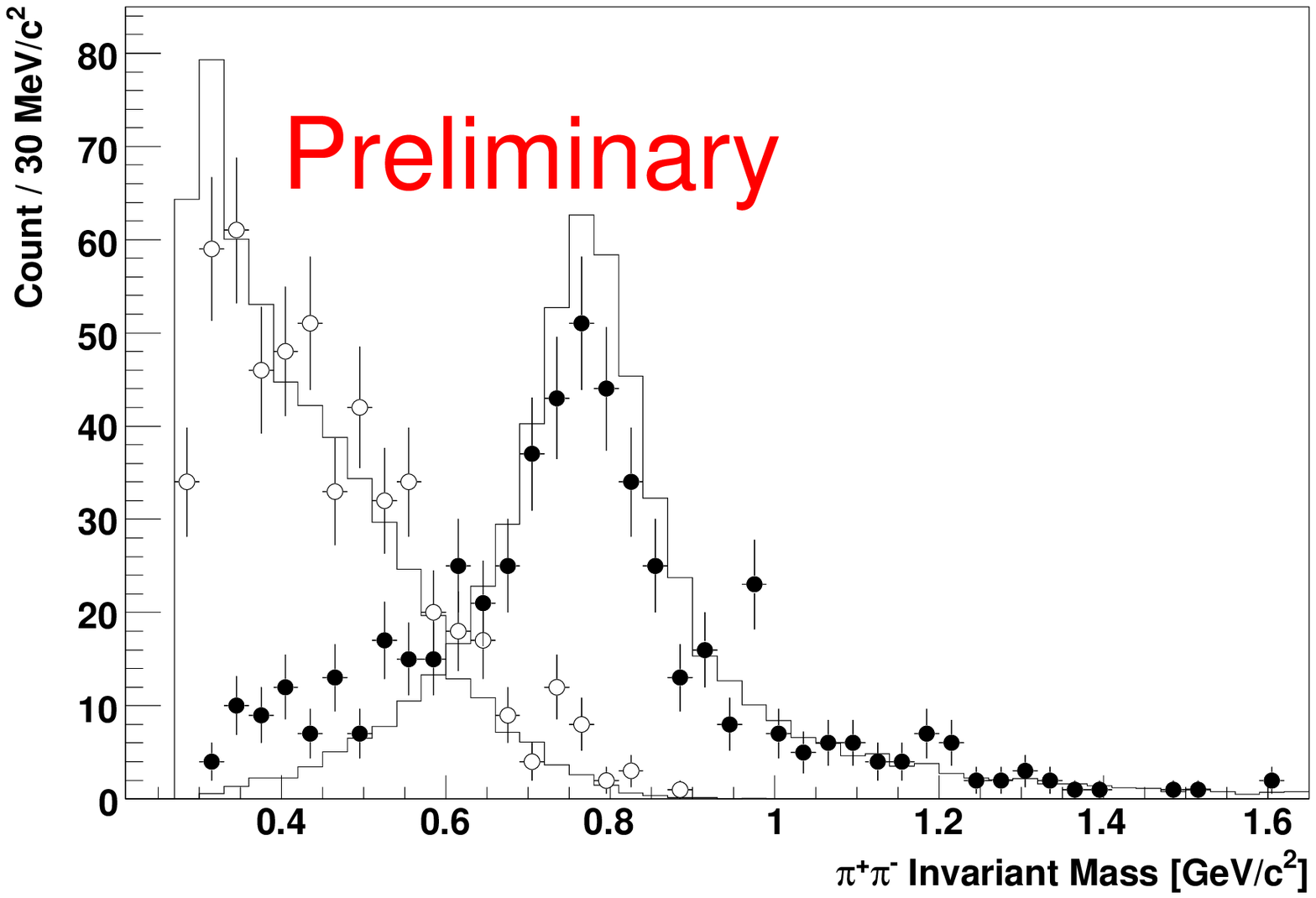} \hfill
  \includegraphics[width=0.49\textwidth]{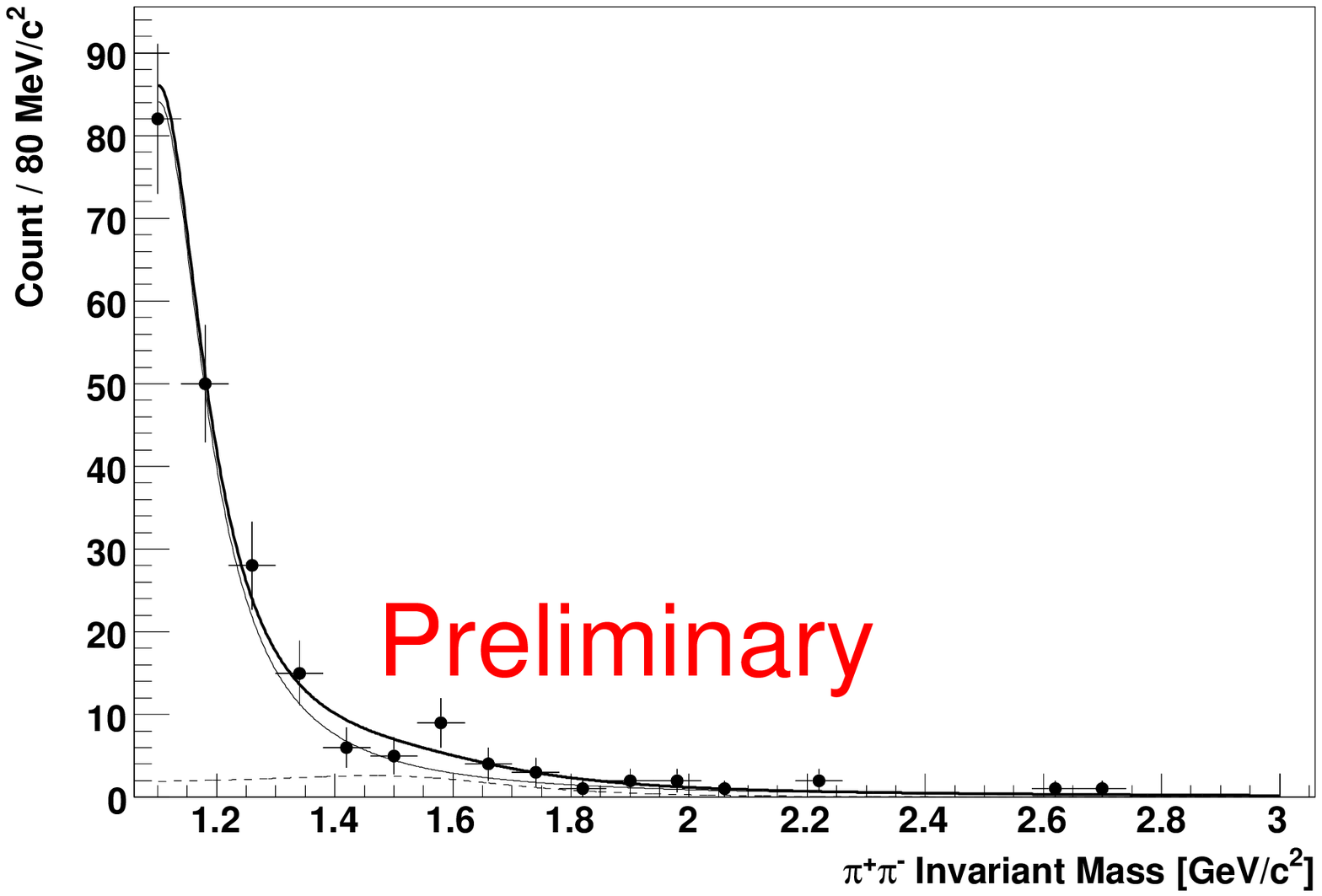} \\
  \raggedright
  \hspace{0.22\textwidth} \textbf{a)} \hspace{0.49\textwidth} \textbf{b)}
  \vspace*{-1ex}
  \caption{a)~Invariant Mass distribution of two-pion subsystems:
    The open circles show the measured invariant mass spectrum of the
    lightest \twopion\ pair in the event. The filled circles represent
    the invariant mass distribution of the \twopion\ that is recoiling
    against the lightest pair. The spectrum exhibits a clear peak in
    the $\Prz(770)$ region. The solid line histograms show the
    prediction from simulation assuming the relative $S$-wave decay
    $\Pr' \to \Prz(770)\,\,\Pf(600)$. b)~High mass region of the
    $m_{\twopion}$ spectrum with tighter cuts applied in order to
    suppress background: The points represent the data. No significant
    enhancement is seen in the region around 1540\mevcc\ where the
    \fourpion\ invariant mass spectrum exhibits a peak.}
  \label{fig:rhoPrimeTwoPi}
\end{figure}

\Figref{rhoPrimeTwoPi}a shows that \fourpion\ events mainly
consist of a low mass \twopion~pair accompanied by a $\Prz(770)$. This
motivated the Monte-Carlo decay model $\Pr' \to \Prz(770)\,\,\Pf(600)$
which is used to estimate the acceptance corrections. As can be seen
in \figref{rhoPrimeTwoPi}a this model reproduces the data well.

In photoproduction on carbon targets the $\Pr'$ was seen not only in
the \fourpion\ decay mode, but also in \twopion\ final
states~\cite{four_pion_fnal}. \Figref{rhoPrimeTwoPi}b shows the high
mass region of the measured $m_{\twopion}$ spectrum. In order to
suppress backgrounds, in particular cosmic rays, tighter cuts are
applied. The data do not show any significant enhancement around
1540\mevcc.

\section{Summary}

STAR has measured photonuclear production of $\Prz(770)$ in
ultra-peripheral relativistic heavy ion collisions. The measured cross
sections agree with model predictions. STAR also measured for the
first time the interference effect in \Prz\ production which indicates
that the decoherence induced by the \Prz\ decay is small and that the
\twopion\ final state wave function is entangled and nonlocal. In
addition STAR has observed coherent photoproduction of \fourpion\
final states in UPCs. The \fourpion\ invariant mass spectrum exhibits
a broad peak around 1540\mevcc\ and no corresponding enhancement is
seen in the $m_{\twopion}$ distribution. The coherent \fourpion\
production cross section is $13.4 \pm 0.8$~\% of that of the
$\Prz(770)$ meson.

\begin{footnotesize}
  
\end{footnotesize}

\end{document}